\documentclass[prl,twocolumn,showpacs,preprintnumbers,amsmath,amssymb]{revtex4}

\usepackage{graphicx}
\usepackage{dcolumn}
\usepackage{bm}

\begin{document}

\title{Superconducting Wigner Vortex Molecule near a Magnetic Disk}

\author{M. V. Milo\v{s}evi\'{c}}
\author{F. M. Peeters}
\email{peeters@uia.ua.ac.be}

\affiliation{Departement Natuurkunde, Universiteit Antwerpen (UIA), \\
Universiteitsplein 1, B-2610 Antwerpen, Belgium}

\date{\today}

\begin{abstract}
Within the non-linear Ginzburg-Landau (GL) theory, we investigate
the vortex structure in a superconducting thin film with a
ferromagnetic disk on top of it. Antivortices are stabilized in
shells around a central core of vortices (or a giant-vortex) with
size-magnetization controlled ``magic numbers''. An equilibrium
vortex phase diagram is constructed. The transition between the
different vortex phases occurs through the creation of a
vortex-antivortex pair under the magnetic disk edge.
\end{abstract}


\maketitle

Almost half a century ago, Abrikosov used the Ginzburg-Landau (GL)
equations to predict that the magnetic field penetrates a type II
superconductor in bundles arranged in a regular lattice, i.e. the
vortex lattice~\cite{abrikosov}. Nowadays, it is well known that
the motion of Abrikosov vortices gives rise to dissipation which
is the limiting factor for the size of the critical current of a
superconductor. The last decade has seen an increased interest in
vortex matter in inhomogeneous superconductors, where defects
(random and ordered) are used to pin the Abrikosov lattice in
order to increase the critical current which is crucial for
practical applications. Irradiation by heavy ions allowed to
create dense random arrays of columnar defects which increased the
critical current density~\cite {givale}. Thanks to substantial
progress in the preparation of magnetic microstructures in
combination with superconductors~\cite{schuller},
superconductor/ferromagnet hybrid systems became very interesting
both theoretically and experimentally. In Ref.~\cite{marm}
Marmorkos {\it et al.} investigated the problem of a magnetic
cylinder with out-of-plane magnetization embedded in a
superconducting film. They solved the non-linear Ginzburg-Landau
equation numerically, with appropriate boundary conditions, and
found a correspondence between the value of the magnetization and
the vorticity of the giant-vortex states.

More recently, the interest shifted to the pinning behavior of
regular arrays of submicron ferromagnetic disks where additional
pinning contributions arise due to the magnetic nature of the
pinning centers~\cite{schuller2,vanbael}. The most prominent
feature of the vortex lattice pinned by a lattice of artificial
defects is the existence of matching fields $H_{n}=n\Phi _{0}/S$,
where $\Phi _{0}$ is the flux quantum, and $S$ is the area of the
primitive cell of the artificial lattice. For matching fields, the
number of vortices per unit cell of the artificial lattice becomes
an integer number. As expected, such a vortex lattice is pinned
much stronger by such artificial defects and the film resistivity
exhibits deep minima~\cite{schuller2} at those matching fields.
These artificial pinning arrays were successfully used to gain
insight into the macroscopic commensurability effects, but the
origin and microscopic nature of this phenomena have not yet been
fully explained. Moreover, the role of the self-magnetic field of
a ferromagnet in the vicinity of the superconductor is not fully
understood.

Therefore, in this letter we consider the interaction between a
single ferromagnetic disk (FD) and a superconducting thin film
(SC), within the non-linear GL theory. The FD lies on top of the
superconductor ($xy$ plane) and it is magnetized in the positive
$z$-direction. To avoid the proximity effect and exchange of
electrons between FD and SC we separate them by a thin layer of
insulating oxide, as is usually the case in experimental
conditions. Moreover, we consider the magnetic disk to be made of
a hard magnet whose uniform magnetic moment and internal currents
are not affected by nearby circulating supercurrents. The
ferromagnetic disk is the only source of magnetic field and we
study in detail how the system is perturbed in the neighborhood of
the disk. The creation of vortices due to this field, as well as
their behavior strongly influence the pinning of additional
external flux lines. Furthermore, we find new ordered
vortex/antivortex structures.

For thin superconductors $(d<\xi ,\lambda )$ it is allowed to
average the GL equations over the film thickness and write them as
\begin{equation}
\left( -i\overrightarrow{\nabla }_{2D}-\overrightarrow{A}\right) ^{2}\Psi
=\Psi \left( 1-\left| \Psi \right| ^{2}\right) ,  \label{lijn1}
\end{equation}
\begin{equation}
-\Delta _{3D}\overrightarrow{A}=\frac{d}{\kappa ^{2}}\delta \left( z\right)
\overrightarrow{j}_{2D},  \label{lijn2}
\end{equation}
where
\begin{equation}
\overrightarrow{j}_{2D}=\frac{1}{2i}\left( \Psi ^{\ast }\overrightarrow{%
\nabla }_{2D}\Psi -\Psi \overrightarrow{\nabla }_{2D}\Psi ^{\ast }\right)
-\left| \Psi \right| ^{2}\overrightarrow{A},  \label{lijn3}
\end{equation}
is the density of superconducting current and $\overrightarrow{A}$
is the total vector potential from the FD and supercurrents, with
boundary condition $\overrightarrow{A}=0$ far away from
\begin{figure*}
\vspace{-0.5cm}
\includegraphics[height=11cm]{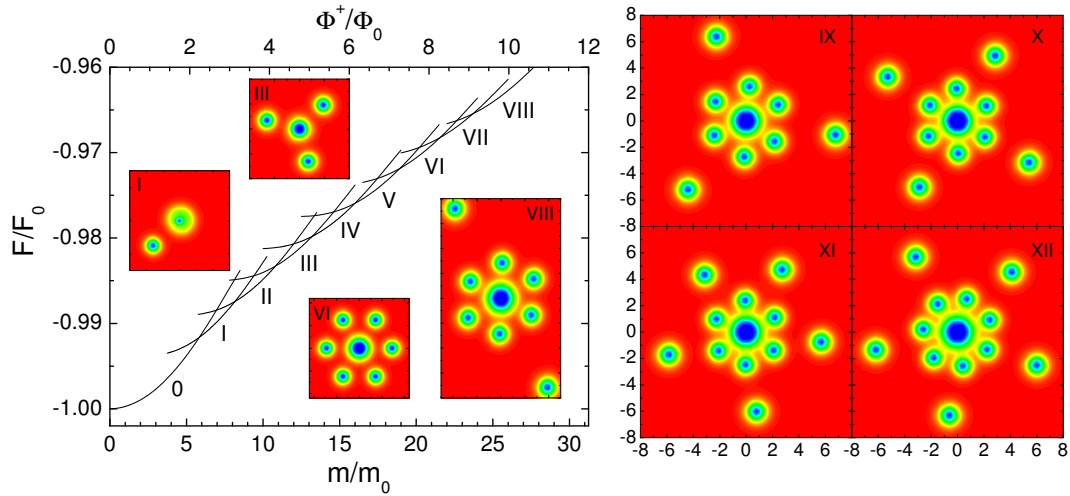}
\vspace{-4.5cm} \caption{\label{fig:fig1}The Gibbs free energy as
a function of the magnetic moment of the disk with radius
$R_{d}/\xi =1$ placed on top of the superconductor. Top axis shows
the flux captured by the superconductor in the positive stray
field area of the magnetic disk. Insets and figures on the right
are the contourplots of the Cooper pair density (red(blue):
high(low) density) for a few ground state vortex configurations. A
giant-vortex is surrounded by antivortices, and the total
vorticity equals zero. Only $1/16$ of the total simulation area is
shown.} \vspace*{-0.3cm}
\end{figure*}
the FD. Here the distance is measured in units of the coherence
length $\xi $, the vector potential in $c\hbar /2e\xi $, and the
magnetic field in $H_{c2}=c\hbar /2e\xi ^{2}=\kappa
\sqrt{2}H_{c}$. The indices $2D$, $3D$ refer to two- and
three-dimensional operators, respectively. To solve the system of
Eqs.~(\ref{lijn1}-\ref{lijn2}), we apply a finite-difference
representation of the order parameter and the vector potential on
a uniform Cartesian space grid (x,y), with typically $512$ grid
points along the width of the rectangular simulation region
$L_{x(y)}$, and use the link variable approach~\cite{kato}, and an
iteration procedure based on the Gauss-Seidel technique to find
$\Psi $. The vector potential is then obtained with the fast
Fourier transform technique. The first GL equation is solved with
an iteration procedure~\cite
{schweigert1}. The dimensionless Gibbs free energy is calculated as $%
F=V^{-1}\int (2(\vec{A}-\vec{A}_{0})\vec{j}_{2D}-|\Psi |^{4})d\vec{r}$,
where integration is performed over the primitive cell volume $V$, and $\vec{%
A}_{0}$ is the vector potential of the magnetic disk. The periodic
boundary conditions for $\vec{A}$ and $\Psi $ have the
form~\cite{doria}
\begin{eqnarray}
\vec{A}(\vec{\rho}+\vec{b}_{i}) &=&\vec{A}(\vec{\rho})+\vec{\nabla}\eta _{i}(%
\vec{\rho}),  \label{per1} \\
\Psi (\vec{\rho}+\vec{b}_{i}) &=&\Psi exp(2\pi i\eta _{i}(\vec{\rho})/\Phi
_{0}),
\end{eqnarray}
where $\vec{b}_{i}$, $i=x,y$ are the lattice vectors, and $\eta _{i}$ is the
gauge potential which cannot be chosen freely but must preserve the single
valuedness of $\vec{A}$ and $\Psi $. These boundary conditions mean that $%
\vec{A}$, $\Psi $ are invariant under lattice translations
combined with specific gauge transformations~\cite{doria}. Other
quantities, such as the magnetic field, the current and the order
parameter density are periodic. We choose $\eta _{x}=A_{0x}$,
$\eta _{y}=A_{0y}$, with the simulation region typically hundred
times larger than the magnetic disk, i.e. $L_{x}>>R_{d}$, which
implies the quantization of the flux, namely total flux through
the superconductor equals zero. These boundary conditions result
in a periodic repetition of not only the superconductor but the
magnetic disk lattice as well. However, due to the large
simulation region, these disks are far from each other and we are
allowed to treat this problem as a single disk on top of an
infinite superconductor.

To find the different vortex configurations, which include the
metastable states, we search for the steady-state solutions of
Eqs.~(\ref{lijn1}) and (\ref{lijn2}) starting from different
randomly generated initial configurations. Then we
increase/decrease slowly the magnetic moment of the magnetic disk
$m$ and recalculate each time the exact vortex structure. We do
this for each vortex configuration in a magnetic moment ($m$)
range where the number of vortices remains constant. By comparing
the Gibbs free energies of the different vortex configurations we
obtain the ground state. The results are shown in
Fig.~\ref{fig:fig1} for the magnetic disk with radius $R_{d}/\xi
=1$ and thickness $d_{d}/\xi =0.5$, on top of the superconductor
with thickness $d/\xi =0.5$. The magnetic moment is expressed in
units of $m_{0}=c\hbar\xi /2e=H_{c2}\xi ^{3}$. Demagnetization
effects are taken into account in this calculation and the
Ginzburg-Landau parameter $\kappa $ was chosen to be $1$, which
approximately corresponds to the experimental values found for Pb,
Nb, or Al films. The values for $d$, $d_{d}$ and $\kappa$
mentioned above will be used throughout this paper.

From Fig.~\ref{fig:fig1}, we notice that with increasing
magnetization of the ferromagnetic disk, the ground state goes
through different vortex states denoted by successive Roman
numbers. In the case of a finite extend of the superconducting
film, i.e. thin mesoscopic disks, with a magnetic disk on
top~\cite{misko}, the Roman numbers corresponded to the vorticity
of the ground state. However, in that case, the total flux
penetrating the finite size superconductor was positive, and
therefore, the states with positive vorticity dominated the free
energy diagram. In the present case, the magnetic disk is on top
of an infinite SC film, consequently all flux is captured by the
superconductor, and the total penetrating flux equals zero. Due to
phase conservation, vortices cannot appear individually, but only
as vortex-antivortex pairs. The vortex is located under the disk,
and the antivortex just outside the positive field region. In
Fig.~\ref{fig:fig1}, each new vortex state corresponds to the
appearance of a new vortex-antivortex pair. Because the magnetic
disk is relatively small, vortices are strongly confined by the
positive magnetic field under the disk, which leads to a
giant-vortex as the energetically favorable state, while
antivortices are symmetrically arranged on a ring around the
central vortex (see Fig.~\ref{fig:fig1}). The Cooper pair density
contourplots in Fig.~\ref{fig:fig1} show the states in question
for the particular value of the magnetic moment of the disk when
they become the ground state, namely at the crossing points of the
free energy curves. With increasing magnetization of the disk and
fixed vortex state, the antivortices move further away from the
magnetic disk. The configurations shown in Fig.~\ref{fig:fig1} are
the result not only of the magnetic field of the ferromagnetic
disk, but also of the interaction between the vortices. While the
giant-vortex strongly attracts the antivortices, the antivortices
repel each other. One therefore expects that the vortex
configurations strongly depend on the size of the magnetic disk,
since it determines the distance between the giant-vortex and the
antivortices, and the circumference of the circle the antivortices
are located on.

Because the vortex configuration is determined by the balance
between the giant vortex-antivortex attraction and the repulsion
between antivortices, it is clear that the number of antivortices
on a ring around the giant-vortex cannot increase monotonously.
Actually, the formation of a second, or of even more rings is
possible with increasing $m$ (see Fig.~\ref{fig:fig1}, for the
states with $N=9-12$, where $N$ denotes the number of
antivortices). A second ring is formed when the number of
antivortices reaches a saturation value $n_{s}$ (or ``magic
number''), in our case this occurs for $n_{s}=6$. This, however,
does not mean that the number of vortices in the first ring
remains constant with increasing magnetization. For example for
$N=12$ it turns out that one more antivortex fits in the first
ring.

\begin{figure}[t]
\vspace{0.5cm}
\includegraphics[height=8cm]{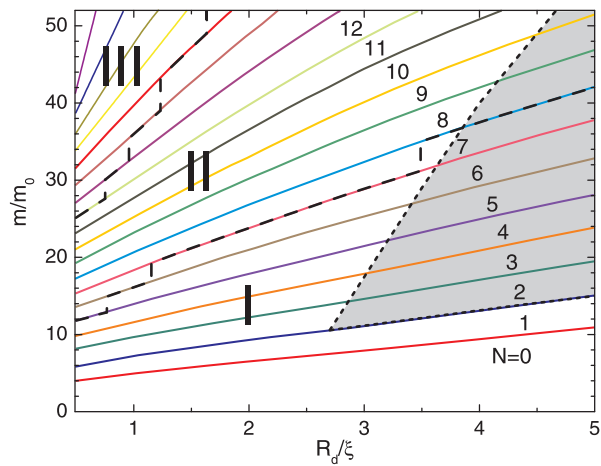}
\vspace{-3cm} \caption{\label{fig:fig2}Phase diagram: the relation
between the radius of the disk, its magnetic moment, the number of
antivortices (N), the number of antivortex rings (Roman numbers)
and the vortex state under the magnetic disk (shaded area:
multivortex state).} \vspace{-0.5cm}
\end{figure}

Since the antivortices are arranged around the positive magnetic
field region, the saturation number $n_{s}$ should depend on the
size of the magnetic disk. Furthermore, for larger disks the wider
area of the positive field region under the magnetic disk could
allow the giant-vortex under the disk to split into individual
vortices. Therefore, we investigated the influence of the disk
size on the vortex structure of the superconducting film.
We enlarged the magnetic disk from $%
R_{d}/\xi =0.5$ to $5$ keeping the thickness $d_{d}$ and the
parameters of the SC fixed. In doing so, we obtain the equilibrium
vortex phase diagram, shown in Fig.~\ref{fig:fig2}. Solid lines
correspond to transitions between the successive $N$ states, and
the thick dashed lines give the number of antivortices needed for
the formation of an additional ring of antivortices. In the white
(shaded) region the giant- (multi-) vortex state is found under
the magnetic disk.

Notice that with enlarging the magnetic disk the number of
antivortices needed before a new ring is formed increases, as
expected. But, on the other hand a larger magnetic moment is
needed for the appearance of a new vortex-antivortex pair in the
ground state. However, the volume of the disk also increases,
which diminishes the
effects of the magnetic moment increase. Therefore, we calculated
the flux $\Phi ^{+}$ through the SC, due to the positive part of
the stray field in the $z=0$ plane, namely in the center of the
superconductor and up to the radius $R_{0}$$\simeq$$R_{d}$ where
the magnetic field changes its polarity. The top axis in
Fig.~\ref{fig:fig1} shows the value of this flux. Such a
calculation tells us that if we keep the magnetic moment constant
and increase the radius (volume) of the disk, the flux $\Phi ^{+}$
decreases, making the positive (and negative) flux through the SC
approximately constant along the transition line between the
successive states in Fig.~\ref{fig:fig2}. Obviously, a larger flux
is needed to create the first vortex-antivortex pair, i.e. $\Delta
\Phi ^{+}/\Phi _{0}=2.112$, due to the high stability of the
Meissner state and the asymmetry of the $N=1$ state (see inset of
Fig.~\ref{fig:fig1}). Further increase of the flux $\Phi ^{+}$
decreases $\Delta \Phi ^{+}$ and for larger $N$ it approaches the
value $\Delta \Phi ^{+}/\Phi _{0}=1.073$. Notice that the flux
$\Phi ^{+}$ is not exactly quantized in units of $\Phi _{0}$ which
is a mesoscopic effect. The quantization condition $\Phi
=\oint_{C}\vec{A}\cdot d\vec{l}=L\Phi _{0}$ cannot be used because
it is not possible to construct a contour $C$ around the positive
stray field region where the current is zero.

\begin{figure}[t]
\vspace{-0.5cm}
\includegraphics[height=10cm]{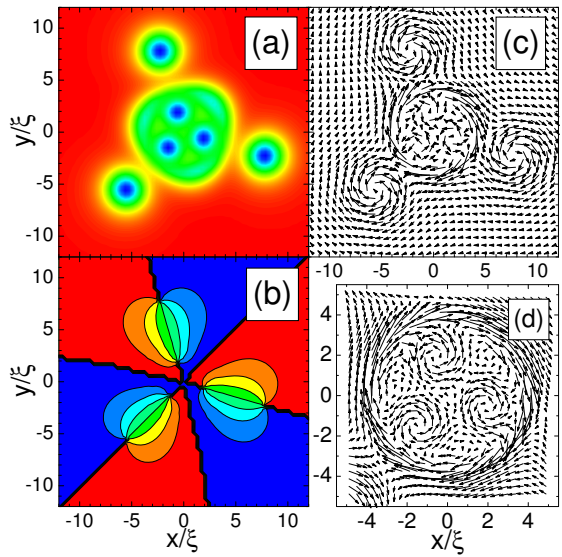}
\vspace{-2.5cm} \caption{\label{fig:fig3}Splitting of the
giant-vortex under the disk into a multivortex: a) Cooper pair
density, b) superconducting phase contourplot (zero - blue,
$2\protect\pi $ - red), for $N=3$, $R_{d}/\protect\xi =4.0$, and
$m/m_{0}=18.5$. (c) is a vectorplot of the corresponding
supercurrents and (d) is an enlargement of the central area.}
\vspace*{-0.3cm}
\end{figure}

With increasing radius of the disk, the giant-vortex splits into
individual vortices, as shown in Fig.~\ref{fig:fig3}(a).
Figure \ref{fig:fig3}(b) shows the phase of the superconducting
order parameter, from which the location of the vortices and
antivortices can be easily deduced. With increasing magnetic
field, these vortices move closer together, resulting ultimately
in the formation of a giant-vortex. Figs.~\ref{fig:fig3}(c,d) show
the current distribution in the SC: the supercurrents around each
vortex and antivortex flow in opposite directions, which are
determined by the direction of the change of the phase of the
order parameter. However, the current around the central vortices
(i.e. at the edge of the magnetic disk) has the same direction as
the one around the antivortices, although the phase change implies
otherwise. The reason is that the direction of this supercurrent
is governed by the vector potential of the magnetic field of the
ferromagnetic disk which is maximal near the edge of the disk. The
value of the total current will be maximal in the same region.
This is also the reason why with increasing magnetic moment of the
FD new vortices nucleate exactly under the edge of the disk, i.e.
when the maximal current reaches the value of the GL current.

\begin{figure}[t]
\includegraphics[height=8.5cm]{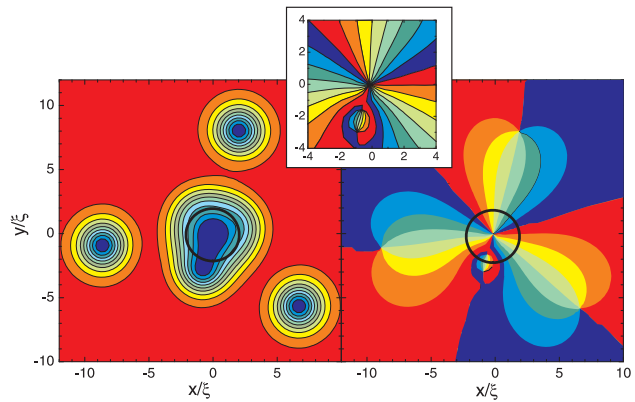}
\vspace{-3.7cm}\caption{\label{fig:fig4}Generation of the
vortex-antivortex pair with increasing magnetic moment of the
disk: a) Cooper pair density, and b) superconducting phase
contourplot, for $N=3(4)$ and $R_{d}/\protect\xi =2.0$. The thick
ring shows the edge of the FD. The upper inset is an enlargement
of the phase of the central area which more clearly shows the
vortex-antivortex pair.} \vspace*{-0.3cm}
\end{figure}

To analyze the generation of new vortices, we start from one
specific initial condition, for example, the ground state
configuration for $N=3$. Then we increase slowly the magnetic
moment of the magnetic disk $m$ and recalculate the vortex
structure within $10^{5}$ iteration steps for each value of $m$.
This way we can find the magnetic moment for which the initial
state changes from $N=3$ to the $N=4$ configuration and we observe
the nucleation of a vortex-antivortex pair by plotting the
intermediate results during the different iteration steps. The
result for the III$\rightarrow$IV transition is shown in
Fig.~\ref{fig:fig4}. The Cooper pair density plot
Fig.~\ref{fig:fig4}(a) shows only a distorted central vortex, but
the superconducting phase plot (b) reveals the real vortex
structure. Namely, in addition to the configuration
similar to the state III, one can notice a vortex-antivortex pair
nucleating under the edge of the dot. Further iterations in our
calculation show that the vortex from this pair will move towards
the central giant-vortex and the antivortex moves to the
periphery, causing the rearrangement of the shell of antivortices
into the configuration IV.

The predicted new vortex configurations can be observed
experimentally by using e.g. scanning probe techniques like Hall
and Magnetic Force Microscopy. Furthermore, these vortex
structures will influence the pinning properties and the
superconducting phase diagram~\cite{lange}.

The authors acknowledge D.~Vodolazov for fruitful discussions.
This work was supported by the Flemish Science Foundation
(FWO-Vl), IUAP, GOA, and the ESF programme on ``Vortex matter''.

\end{document}